\documentclass[useAMS,usenatbib]{mn2e}
\usepackage{epsfig}
\usepackage{graphicx}
\usepackage{epstopdf}
\newcommand\mg{{M_{\rm g}}}

\def\simgt{\lower.5ex\hbox{$\; \buildrel > \over \sim \;$}}
\def\simlt{\lower.5ex\hbox{$\; \buildrel < \over \sim \;$}}

\def\c12{$^{12}$C}
\def\neo20{$^{20}$Ne}
\def\al27{$^{27}$Al}
\def\ne22{$^{22}$Ne}
\def\na23{$^{23}$Na}
\def\mg25{$^{25}$Mg}
\def\mag26{$^{26}$Mg}

\title[Dilution constraints in GCs]{Formation of Multiple Populations
  in Globular Clusters: constraints on the dilution by pristine gas}

\author[A. D'Ercole et al.]{Annibale D'Ercole$^1$\thanks{E-mail:
    annibale.dercole@oabo.inaf.it}, Francesca D'Antona$^2$, Enrico
  Vesperini$^3$\\$^{1}$INAF- Osservatorio Astronomico di Bologna, via
  Ranzani 1, I-40127 BOLOGNA (Italy)\\$^{2}$INAF- Osservatorio
  Astronomico di Roma, via di Frascati 33, I-00040 Monteporzio
  (Italy)\\$^3$Department of Physics, Drexel University, Philadelphia,
  PA 19104, USA}

\begin{document}

\date{Accepted ... Received ...; in original form ...}

\pagerange{\pageref{firstpage}--\pageref{lastpage}} \pubyear{2002}

\maketitle

\label{firstpage}

\begin{abstract}  
  The star-to-star differences in the abundance of light elements
  observed in the globular clusters (GCs) can be explained assuming
  that a second generation (SG) of stars form in the gas ejected by
  the asymptotic giant branch (AGB) stars belonging to a first stellar
  generation. However, while Na and O appear to be anticorrelated in
  the cluster stars, from the stellar models they turn out to be
  correlated into the AGB ejecta. In order to reconcile the stellar
  theory with the observational findings, all the GC models invoke an
  early dilution of AGB ejecta with pristine gas occurring during the
  SG formation.  Despite a vast consensus about the occurrence of such
  dilution, the physical process behind it is still unknown. In the
  present paper we set some general constraints on the pristine gas
  dynamics and on the possible amount of pristine gas involved in the
  SG formation, making use of a one zone chemical model. We find that
  such a dilution is a necessary ingredient in the SG star formation
  to explain the observed abundance patterns. We confirm the
  conclusion of our previous works showing that clusters must have
  been initially much more massive. We also show that models assuming
  that clusters had an initial mass similar to their current one, and
  adopting a large fraction of pristine gas to form SG stars, fail to
  reproduce the observed Na-O anticorrelation and are not viable.  We
  finally show that the dilution event should be restricted in time,
  rather than extended for the all duration of the SG formation.
\end{abstract}


\section{Introduction} 
\label{sec:introduction}

In the last decade a number of spectroscopic data provided the evidence
of spreads of light elements and anticorrelations between Na and O and
Mg and Al in the stars of globular clusters (GCs), indicating the
presence of at least two stellar populations in these objects. As these
properties are not shared by the halo stars with the same metallicity,
nor by open clusters in our or other galaxies, they represent a trait
of the GCs so peculiar to lead \citet{carr10} to identify GCs as those
clusters where there is a Na-O anticorrelation. On the basis of this
anticorrelation, \citet{carr10} separate the GC stars in three
families: a Primordial population of stars sharing the same abundances
of the halo stars; an Intermediate population of stars with a lower O
abundance and higher Na; and an Extreme population of very O-poor stars
([O/Fe]$<-0.4$). 

The variations are also found among unevolved stars
currently on the main sequence (MS). This implies that the above
chemical characteristics have been imprinted in the gas by a
previous generation of stars, because low-mass MS stars do not reach
the high temperatures needed to sustain the nuclear reactions leading
to the observed relations between the elements. 

It is still debated
which kind of stars act as polluters, whether asymptotic giant branch
(AGB) stars \citep{dantonaventura2007} or fast rotating massive stars
\citep[FRMS; ][]{decressin2007}. In \citet{der08}, we presented a model
for the formation and dynamical evolution of multiple populations in
GCs in which second generation (SG) stars form out of the AGB ejecta of
the first generation (FG) stars. We first focussed on the hydrodynamic
of the ejecta that collect into the cluster core through a cooling
flow; it turned out that, in order to shed enough gas to form the
observed SG, the FG should have been originally about ten times larger
than that observed today. By means of N-body simulations we then showed
how the cluster can lose FG stars reaching the nowadays observed ratio
of the number of FG and SG stars. 

In a successive paper \citep[][hereafter Paper I]{der10}, we focused
on the chemical evolution of the GCs through simple one-zone models,
and we were able to reproduce the main characteristics of clusters
such as NGC 2808 and M4. The main problem encountered in this kind of
models is given by the fact that the current stellar models indicate
that during the AGB phase [O/Fe] and [Na/Fe] in the ejecta are always
correlated
\citep[e.g.][]{karakas2007,ventura2009,siess10,ventura2010}. Thus, the
O and Na abundances in the cluster interstellar medium (ISM) are
correlated, and the SG stars forming in this gas would exhibit the
same correlation, in glaring conflict with the observations.  Also in
the FRMS scenario some degree of dilution with pristine gas is needed
to account for the presence of lithium in SG stars because this
fragile element is absent in the massive star ejecta
\citep{decressin07b,lind11}.  To date, dilution seems to be required by all
the proposed models.

Dilution was actually
implemented in the models presented in Paper I. This implementation,
however, was realized via a reasonable but arbitrary modelling. In
fact, despite the general consensus about the need of dilution, the
source and dynamics of the diluting pristine gas is still not clear,
and the mechanisms proposed till
now \citep{der08,bekmac09,pflkro09,conspe10,gratcar10} all suffer
from some drawbacks. 

The aim of the present paper is to shed some light
on the dilution process through possible constraints that can be
inferred by the chemical properties of the GCs.

\section{On the need of dilution}
\label{sec:need}

Given the vagueness of the dilution process and the great uncertainties
plaguing some relevant aspects of the nuclear physics during the AGB
stage \citep[][Paper I]{ventura2005a,ventura2005b},one could be
tempted to make the simple (but audacious) statement that the actual
stellar models are incorrect, and that the Na abundance in the AGB
atmosphere decreases in pace with the decrease of stellar mass. In
the end, the O-Na anticorrelation shown by the SG stars would be
the direct consequence of the O-Na anticorrelation already present in
the gas in which these stars form. 

Unfortunately, although the above
picture eliminates some difficulties connected with the dilution, other
problems remain unsolved. While the O-Na anticorrelation seems to be a
distinctive characteristic of GCs, the extension of this
anticorrelation varies from cluster to cluster
\citep{carretta2009a}. In the previous scenario the variation of such
extension -- that is, of the minimum value [O/Fe]$_{\rm min}$
\footnote{The extension of the anticorrelation is regulated by the
  value of [O/Fe]$_{\rm min}$ as the maximum value is essentially
  constant in all the clusters and is the same found in the halo
  stars.} -- depends on the time at which the star formation of the SG
starts. For the sake of example, we can examine a ``fake'' evolution of
[Na/Fe] within the gas shed by the AGB stars, with the Na abundance
decreasing in time and the oxygen which instead grows. Let us now
consider two clusters, GC\#1 and GC\#2, starting to form SG stars at
the times $t_{\rm 1}$ and $t_{\rm 2}>t_{\rm 1}$, respectively. Clearly,
GC\#1 will host some very O-poor stars which instead will be absent in
GC\#2, and therefore the anticorrelation in this latter cluster will be
shorter.  On the other hand, given our assumption about the Na
evolution, the maximum value [Na/Fe]$_{\rm max}$ will be larger in
GC\#1 than in GC\#2. Thus, an anticorrelation is expected between
[O/Fe]$_{\rm min}$ and[Na/Fe]$_{\rm max}$, at odd with the
observations showing instead a correlation \citep[see Fig. 19
in][]{carretta2009a}.

While this correlation is not quantitatively
accounted for by any model yet, we think that in any case it falsifies
the starting hypothesis that the anticorrelation O--Na is imprinted in
the stellar models. Therefore, despite the problems outlined above, we
more safely acknowledge the current models of stellar evolution, and
believe that the dilution is an inescapable ingredient of the GC
evolution.

\section{Constraints on the FG mass}
\label{sec:fgmas}

As discussed in \citet{der08}, in order to explain the current mass
of SG stars, globular clusters must have been initially about ten
times larger than today\footnote{The actual factor depends on the
  stellar IMF, the duration of the SG formation event, the amount of
  ejecta a cluster is able to retain
  \citep[see][]{ves10}}. In \citet{ves10}, we have further explored the
actual amount of SG stars that a cluster could have formed depending on
its initial mass and structural parameters (along with the implications
for the contribution of globular cluster FG and SG stars to the
assembly of the Galactic halo).

However, \citet{decressin07b} leave open
the possibility, in the FRMS scenario, of an initial FG mass close to
the present one, provided a rather flat IMF. The same hypothesis was
put forward by \citet{daca04} assuming the AGB stars as polluters.  As
shown by \citet{der08}, a flatter IMF increases the amount of available
ejecta to form the SG population at the cost of reducing the long-lived
FG stars, and a large $M_{\rm FG}$ is still needed to raise their
number; as a consequence, a flat IMF is not a viable hypothesis to
avoid larger clusters in the past. 

More recently also \citet{conspe10}
have proposed a model, still based on the AGB scenario, in which the
cluster has an initial mass similar to the current one; the SG stars
form from the small amount of AGB ejecta available and a very large
amount of pristine gas accreted by the GC while moving through the
ambient ISM .  This last scenario can be easily ruled out for those few
clusters in which a very high helium enrichment ($Y\sim 0.38$) has been
safely derived from their CMDs because such a large amount of helium
can not be released by a number of FG stars as small as the present one
\citep{ren08}.

Here we show that, apart from the helium abundance, the
assumption of a FG comparable in mass to the present one looks
impractical on the basis of the observed anticorrelation between O and
Na. Let us assume, as in Paper I, a SG star formation starting at
$t=32$ Myr and lasting up to a fiducial time $t_{\rm end}=100$
Myr. During this lapse of time the FG of mass $M_{\rm FG}$ with a
Kroupa IMF \citep{kroupa93} sheds an amount of ejecta $M_{\rm ej}=0.05
M_{\rm FG}$ \citep[see][]{der08}. If we now further assume to form an
amount of SG stars comparable to the FG, as suggested by
\citet{conspe10}, the quantity of the accreted pristine gas must be of
the order of $M_{\rm pr}\sim M_{\rm FG}$. In conclusion, the degree of
pollution of the SG is about $M_{\rm ej}/M_{\rm pr}=0.05$, which
is definitively too low to explain the observed extension of the
O-Na anticorrelation \footnote{Note that we have minimized the dilution
  posing $M_{\rm SG}/M_{\rm FG}=1$; assuming a larger ratio $M_{\rm
    SG}/M_{\rm FG}$ (as usually observed) would lead to an even more
  irrelevant pollution.}.

\begin{figure}    
\centering{\includegraphics[width=8cm]{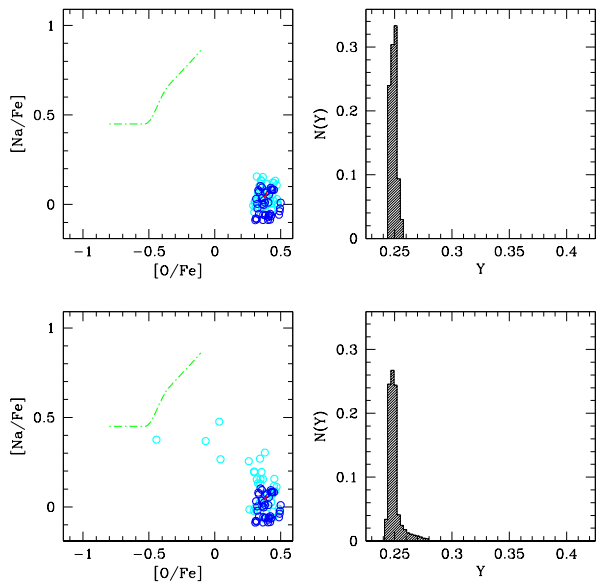}}
\caption{Model with uniform accretion with the following
  parameters:$(t_{\rm end,7},\rho_{\rm *,FG},f_{\rm
    pr},\nu,x)=(10,700,1,1,0.5)$ Top-left panel: chemical evolution of
  the homogeneous model in the plane [O/Fe]-[Na/Fe]. In this panel the
  dot-dashed line represents the ISM chemical pattern in absence of
  any accretion of pristine gas, and the dashed line (barely visible)
  the actual GC evolution for the assumed dilution. The FG stars are
  represented by the blue circles, and the SG stars by the cyan
  circles. Top-right panel: helium distribution of the homogeneous
  model. Bottom-left and bottom-right panels are similar to the upper
  ones, but in case of inhomogeneous mixing.  }
\label{fig:ceschm} 
\end{figure}

The above conclusion is highlighted by our
chemical one-zone model shown in in Fig. \ref{fig:ceschm} .  Adopting
the formalism of Paper I, the model is characterized by the following
set of parameters:$(t_{\rm end,7},\rho_{\rm *,FG},f_{\rm
  pr},\nu,x)=(10,700,1,1,0.5)$. Here $t_{\rm end,7}$ represents
the time at which the evolution stops in units of $10^7$ yr, $\rho_{\rm
  *,FG}$ is the FG density in M$_{\odot}$ pc$^{-3}$, $f_{\rm pr}$ is
the amount of accreted gas in unit of $\rho_{\rm *,FG}$, $\nu$ is
the star formation efficiency, and $x$ is the ratio between SG and
FG stars. Contrary to the models shown in Paper I, where the
accretion rate of the pristine gas is assumed to have a gaussian
temporal profile, here we set a constant accretion rate, as expected in
the case envisaged by \citet{conspe10} of a GC moving with
constant velocity through an uniform medium. 

The top panels of figure
\ref{fig:ceschm} show that the FG and SG stars occupy the same region
in the [Na/Fe]-[O/Fe] diagram and share the same helium distribution
extremely peaked around $Y=0.246$, the value assumed for the pristine
gas and the FG stars. The two populations are thus
undistinguishable. 

Inhomogeneous mixing has been suggested to overcome
the above shortcoming and to obtain the observed
anticorrelations \citep{conspe10}. However, a simple argument shows
that resorting to inhomogeneous mixing does not solve this problem. Let
us consider an amount of AGB ejecta $M_{\rm ej}$ with a typical sodium
abundance [Na/Fe]=0.8 merging with a mass $M_{\rm pr}$ of pristine gas
with, for example, [Na/Fe]=0. In case of inhomogeneous mixing, different
regions of the mixture would have different values of [Na/Fe], and the
stars forming in this gas would have a distribution function
$N$([Na/Fe]) extending across the range $0<$[Na/Fe]$<0.8$. Although the
shape of this function depends on a number of assumptions (see the
Appendix), some conclusions can be drawn independently of its exact
knowledge. \citet{carr10} separate the FG from the SG on the basis of
their sodium abundance, with the SG stars
having [Na/Fe]$>0.4$\footnote{Actually, the separating value is not the
  same for all the clusters, but is 0.4 dex larger than the minimum
  value of [Na/Fe] observed in the cluster.}.  Thus, assuming that all
the mass $M_{\rm ej}$ of the AGB ejecta forms a number $N_{\rm tot}$
of stars, only a fraction $\zeta=N(\rm {[Na/Fe]}>0.4)/N_{\rm tot}$
of them will appear as SG stars, the rest remaining confused with
the genuine FG stars.  The apparent ratio between first and
second generation therefore will be:

\begin{equation}
\label{eq:ratio}
{M_{\rm SG} \over M_{\rm FG}}={M_{\rm SG, {[Na/Fe]}>0.4}\over M_{\rm FG}+M_{\rm SG, {[Na/Fe]}<0.4}}={\zeta \xi \over{1+(1-\zeta)\xi}}. 
\end{equation}
\noindent
In the above equation we have assumed that the mass lost by the FG
stars (before $t=100$ Myr) is$M_{\rm ej}=\xi M_{\rm FG}$. From the
observations we know that $M_{\rm SG}/M_{\rm FG}\ga 1$, and therefore
equation \ref{eq:ratio} gives rise to the condition

\begin{equation}
\label{eq:ratiob}
0.5{1+\xi\over{\xi}}\la \zeta \la 1. 
\end{equation}
\noindent

This inequality is fulfilled only for $\xi \ga 1$, which is
clearly meaningless \footnote{In models in which the FG is at least ten
 times more massive than the SG, the above condition becomes $\zeta
  \la 0.1(1+\xi)/\xi$, and is easily met for any value of $\xi$.}.

In
summary, if the initial FG mass was similar to the present one, the
observed O-Na anticorrelation cannot be reproduced in any way by
inhomogeneous mixing. 

To give a ``visual'' support to the above
conclusion, we have implemented an algorithm in our one-zone code in
order to take into account inhomogeneous mixing (see the Appendix).
The results are shown in the lower panels of Fig. \ref{fig:ceschm}. Non
homogeneous mixing clearly acts in the right direction, but the number
of Intermediate stars turns out to be rather small because of the
large overabundance of pristine gas compared to the AGB
ejecta\footnote{in order to maximize the effect of the inhomogeneity,
 we assumed $\alpha=1.5$, $\alpha$ being a parameter regulating the
 shape of $N$([Na/Fe]) (see the Appendix).}. This is particularly
appreciable in the helium distribution, which acquire only a small tail
toward higher values of $Y$, although the ejecta of massive AGB stars
have values as high as $Y=0.38$. 

We thus conclude that a large FG mass
must be a prerequisite of every model of GC evolution, which then must
account for how to get rid of the stellar surplus \citep[see][]{der08}.

At the end of this section, we mention the effect of a
possible inhomogeneous mixing in models similar to those of Paper I,
with an initial $M_{\rm FG}$ much larger than today, and $x \sim 1$. In
this case it turns out that, in order to create the same amount of
recognizable SG stars, inhomogeneous mixing models require a FG mass
10-70\% larger than in the homogeneous case (see the Appendix).

\section{Constraints on the temporal evolution of dilution }
\label{sec:cons}

In the previous section we considered the case of a globular
cluster moving in the halo through the ambient medium that can be
captured via the Bondi accretion \citep[e.g.][]{limu07}. In addition to
this process, the moving cluster may accrete further material sweeping
up the surrounding gas which impinges on a seed of AGB ejecta present
in the GC centre.  Both the above accretion mechanisms may work only
for a limited range of values of the parameters of interest. In fact,
if the energy of the incoming gas is much larger than the GC potential
well, a collective accretion is prevented. Moreover, if the cluster
velocity is too large, the ram pressure exerted on the gas at the
bottom of the potential well becomes stronger than the gravitational
restoring force, and the sweeping process clears out the cluster of its
gas instead of promoting accretion \citep[see][]{limu07}.

A full
hydrodynamic study of the processes depicted above will be presented in
a separate paper (D'Ercole et al. in preparation). Here we want to
constrain the accretion process on the basis of chemical arguments. If
GCs actually accrete gas from the ambient medium continuously since the
beginning, as suggested by \citet{conspe10}, dilution starts quite soon
and the formation of the Extreme population is inhibited because the
stars of this population are extremely O-poor, and therefore are
expected to be formed in the pure ejecta of the massive AGB stars.

\begin{figure}    
\centering{\includegraphics[width=8cm]{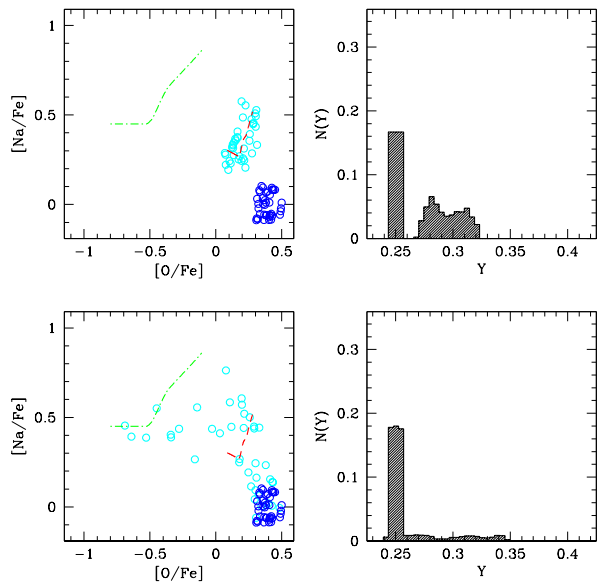}}
\caption{Model similar to that illustrated in figure \ref{fig:ceschm},
  but with $f_{\rm pr}=0.05$. Top panels: homogeneous
  case. Bottom panels: inhomogeneous case.  }
\label{fig:ram} 
\end{figure}

In order to explicate this point, we run a model similar to that
shown in the previous section, but assuming an amount of accreted
matter which is only 5\% of the FG mass, a characteristic value in the
models shown in Paper I: we thus assume $(t_{\rm end,7},\rho_{\rm
  *,FG},f_{\rm pr},\nu,x)=(10,700,0.05,1,0.5)$.  As expected, in
the homogeneous mixing case Fig. \ref{fig:ram} (upper panel) shows
that the stars have all [O/Fe]$>$0. Once the inhomogeneous mixing
is allowed (see Fig. \ref{fig:ram}, lower panel), some O-poor stars
actually form, but they remain too few to fit the amount of the Extreme
(and Intermediate) stars in the GC where they are observed,
reaching 15-25\% of the total \citep{dc2008}. This is well illustrated by the
helium distribution function which barely reaches $Y\sim 0.33$ in the
model, while nearly 20\% of the stars in NGC 2808 has
$Y>0.35$ \citep{carretta2009a,dc2008}.  Moreover, comparing the
two distribution functions in Fig. \ref{fig:ram}, we note that in
the inhomogeneous case the peak at $Y\sim 0.25$ given by the FG stars
is $higher$, reducing the ratio $M_{\rm SG}/M_{\rm FG}$ with respect
to the homogeneous case. This is an effect due to the spread of the
SG population which partially overlaps the Primordial stars (see
the discussion in the Appendix).

The above argument rules out not only
continuous accretion of external pristine gas, but also the $internal$
continuous dilution suggested by \citet{gratcar10} with the pristine
gas produced by mass lost from less evolved stars (typically - but not
only - MS stars) than those producing the polluting material. 

We thus
conclude that the dilution can not be achieved by a continuous mixing
of pristine gas, at least for those clusters showing the presence of an
Extreme population.  We remind here that the models shown in Paper I
may form O-poor stars because the bulk of the accretion occurs at a
time $t_{\rm ac}$ larger than the time at which the massive AGB stars
start to pollute the cluster. Therefore, accretion seems to be a quite
episodic event, rather than a regular one.

\section{Conclusions}

In this paper we highlighted some characteristics of the
dilution process occurring in the CGs on the basis of their
chemical properties. Our findings can be summarized as follows:

\begin{enumerate}
\item In principle, one could reproduce the ubiquitous O-Na
  anticorrelation present in the GCs without any dilution, simply
  assuming that all the available AGB models are incorrect and that
  actually the ejecta of the smaller AGB stars are less abundant in
  sodium and more abundant in oxygen.  However, we have shown that
  this option appears to be ruled out by the direct correlation
  between [O/Fe]$_{\rm min}$-[Na/Fe]$_{\rm max}$ found by
  \citet{carretta2009a}.  We thus conclude that dilution must have
  actually occurred in GCs.
\item We have studied the dependence of the extent of the Na-O
  anticorrelation on the amount of pristine gas involved in the SG
  formation. Our analysis shows that the contribution of pristine gas
  to the matter from which SG stars form can not be larger than $\sim
  0.1M_{\rm FG}$.  This confirms one of the conclusions of our
  previous papers \citep[Paper I;][]{der08,ves10}: in order to form
  the amount of SG stars observed today, clusters must have been
  initially about ten times more massive. Models suggesting an initial
  mass similar to the present one, and based on a flat ISM
  \citep{daca04,decressin07b}, have been ruled out by \citet{der08}.
  A model recently proposed by \citet{conspe10} suggests that
  clusters would not need to be initially much more massive than today
  since SG stars would form almost entirely from a large amount of
  accreted pristine gas.  Following \citet{ren08}, such a model can be
  easily ruled out for those massive clusters containing a very helium
  rich population, because a cluster with a FG mass initially similar
  to the present one can not release the needed helium mass. The
  results presented in this paper imply that the model of
  \citet{conspe10} is ruled out also for all the other clusters
  without any Extreme Helium-rich population. In fact, if the accreted
  mass is of the same order of the FG, the abundances of the SG stars
  are almost identical to those of the FG stars, thus preventing the
  formation of the O-Na anticorrelation which instead is present in
  all GCs. Even a possible inhomogeneous mixing between the AGB ejecta
  and the accreted pristine gas, as invoked by \citet{conspe10}, can
  not give rise to the Extreme SG population and to a substantial
  Intermediate SG population.
\item We demonstrated that if the dilution starts at the same time of
  the pollution, the Extreme stars can not form, because such O-poor
  stars are presumably built in the pure ejecta of massive AGB
  stars. Thus, some proposed accretion processes such as the Bondi
  accretion and streaming \citep{conspe10}, or the self-dilution due
  to unevolved stars \citep{gratcar10}, can not work, at least for
  clusters hosting an Extreme population. Instead, dilution should be
  considered as a delayed episode restricted in time, as described in
  Paper I.
\end{enumerate}

\section*{Acknowledgments}
AD and FD were supported in part by INAF
under PRIN 2009. EV was supported in part by NASA grant NNX10AD86G.

\bibliographystyle{mn2e} 
\bibliography{gc_ref}
%\label{lastpage}

\appendix
\section{Inhomogeneous mixing}
\label{sec:app}

In order to account for a possible inhomogeneous mixing between
the pristine gas and the AGB ejecta we adopt the following scheme. We
assume that both the pristine gas and the ejecta are clumpy, and the
number of these clumps or clouds is distributed in mass following a
power law:

\begin{equation}
\label{eq:np}
{dM_{\rm pr} \over dM}=AM^{1-\alpha}
\end{equation}
\begin{equation}
\label{eq:ne}
{dM_{\rm ej} \over dM}=BM^{1-\beta},
\end{equation}
\noindent
where $dM_{\rm pr}$ ($dM_{\rm ej}$) is the mass of pristine gas
(ejecta) included in clouds with masses between $M$ and $M+dM$. As a
consequence, the total mass $M_{\rm pr,tot}$ of pristine gas, and
$M_{\rm ej,tot}$ of the ejecta, are given by

\begin{equation}
\label{eq:mpt}
M_{\rm pr,tot}={A \over {2-\alpha}}\left (M_{\rm pr,u}^{2-\alpha}-M_{\rm pr,l}^{2-\alpha}\right ),
\end{equation}
\begin{equation}
\label{eq:npt}
M_{\rm ej,tot}={B \over {2-\beta}}\left (M_{\rm ej,u}^{2-\beta}-M_{\rm ej,l}^{2-\beta}\right )
\end{equation}
\noindent
where $M_{\rm pr,u}$ ($M_{\rm ej,u}$) and $M_{\rm pr,l}$ ($M_{\rm
  ej,l}$)are the lower and upper extremes of the pristine gas
(ejecta) distribution, and $A$ and $B$ are normalization constants. As
in our program we form stars as a sequel of ``bursts'' occurring after
every time interval $\Delta t$, $M_{\rm pr,tot}$ and $M_{\rm ej,tot}$
represent the amount of pristine gas and ejecta present during $\Delta
t$, and are given by the simulation. As a consequence, taking $M_{\rm
  pr,u}$, $M_{\rm ej,u}$, $M_{\rm pr,l}$, $M_{\rm ej,l}$, $\alpha$ and
$\beta$ as free parameters, $A$ and $B$ are determined.  Choosing an
uniform deviate $m_{\rm pr}$ in the range $0<m_{\rm pr}<M_{\rm pr,tot}$
we generate a random mass $M_{\rm pr}$ of pristine cloud from the
distribution in eq. \ref{eq:np}, and we proceed similarly to obtain a
random value of the ejecta cloud $M_{\rm ej}$ \citep[e.g.][]{numrec92}:

\begin{equation}
\label{eq:mpran}M_{\rm pr}=\left ({2-\alpha \over {A}}m_{\rm  pr}+M_{\rm pr,l}^{2-\alpha}\right )^{1/(2-\alpha)}
\end{equation}
\begin{equation}
\label{eq:meran}
M_{\rm ej}=\left ({2-\beta \over {B}}m_{\rm  ej}+M_{\rm ej,l}^{2-\beta}\right )^{1/(2-\beta)}.
\end{equation}

Assuming that these two clouds merge together, the resulting
abundance ${\cal Z}$ in this mixed gas is given by

\begin{equation}
\label{eq:met}
{\cal Z}={{\cal Z}_{\rm pr} M_{\rm pr}+{\cal Z}_{\rm ej}M_{\rm ej} \over{M_{\rm  pr}+M_{\rm ej}}},
\end{equation}
\noindent
where ${\cal Z}_{\rm pr}$ and ${\cal Z}_{\rm ej}$ represent the
abundance of the element in the pristine gas and in the ejecta,
respectively. 

After every $\Delta t$ we chose a number of couples
$(M_{\rm pr},M_{\rm ej})$ obtaining the ${\cal Z}$-distribution of the
stars forming at that time. 

As pointed out above, we have six free
parameters that must be fixed. For illustration, we assume $M_{\rm
  pr,u}=10^4M_{\rm pr,l}$, $M_{\rm pr,l}=1$ M$_{\odot}$. For the sake
of simplicity, we further assume $\beta=\alpha$, $M_{\rm ej,u}/M_{\rm
  ej,l}=M_{\rm pr,u}/M_{\rm pr,l}$, and $N_{\rm pr,tot}=N_{\rm
  ej,tot}$ (i.e. the number of pristine clouds is equal to the clouds
of the ejecta), thus obtaining $M_{\rm ej,l}/M_{\rm pr,l}=M_{\rm
  ej,tot}/M_{\rm pr,tot}=x$ which is given by the simulation at every
$\Delta t$.

\begin{figure}    
\centering{\includegraphics[width=8cm]{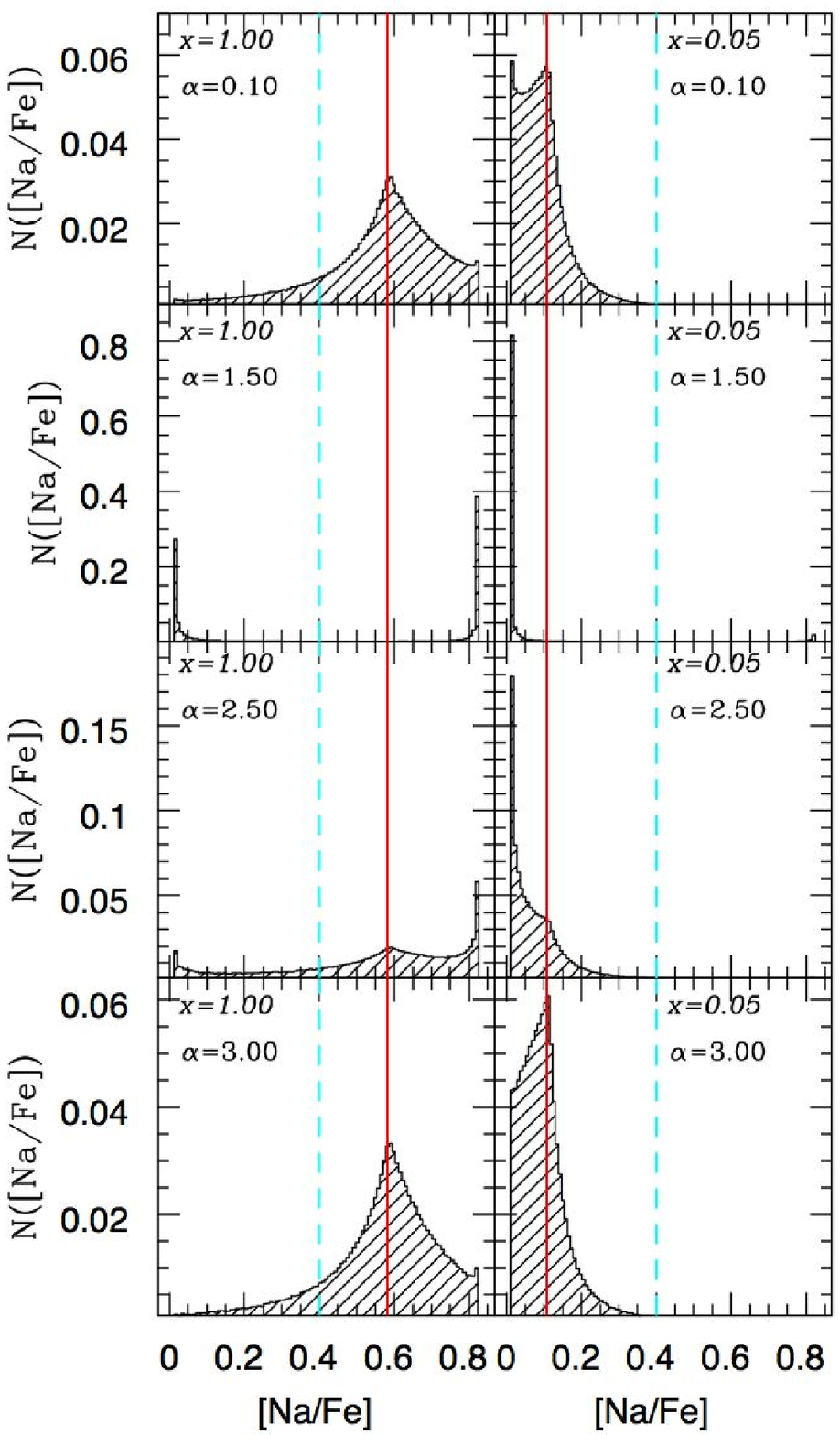}}
\caption{Different distributions of $N$([Na/Fe]) for different values
  of$\alpha$ and $x$. The vertical solid lines indicate the mean
  value in case of homogeneous mixing ($N$([Na/Fe]) would be shaped as
  a delta dirac function). The dashed lines separate the FG population
  ([Na/Fe]$<0.4$) from the SG population
  ([Na/Fe]$>0.4$)\citep{carr10}. For $x=0.05$ the inhomogeneous mixing
  is notable to populate the SG region for any value of $\alpha$.  }
\label{fig:distributions} 
\end{figure}

In order to show the behaviour of the distribution for different values
of $\alpha$, Fig. \ref{fig:distributions} displays several histograms
of the number of stars $N$([Na/Fe]) as a function of [Na/Fe] forming in
a mixture of two gases, one with [Na/Fe]=0,representing the pristine
gas, and one with [Na/Fe]=0.82, representing the AGB ejecta. Two ratios
between the masses of these two gases are taken into account: $x=1$,
characteristic of models like those in Paper I, and $x=0.05$, typical
of models as those suggested by \citet{conspe10}, in which the original
FG mass is the same observed today. 

We note that very steep ($\alpha=3$)
and very flat ($\alpha=0.1$) mass functions give rise to similar
distributions. For low values of $\alpha$ the number of small clouds
and massive clouds are similar, thus the probability of encounters
between clouds with similar mass is elevated, and the final stellar
distribution peaks around the mean value which would be obtained in
case of homogeneous mixing (vertical solid lines in
Fig. \ref{fig:distributions}). In case of high values of $\alpha$, the
majority of the mass is distributed in small clouds of similar mass;
merging between these clouds are the most probable, and once more the
final distribution peaks at the mean value. For intermediate values of
$\alpha$, the larger clouds compensate their lower number with their
larger mass. When they make the most probable merging with a small
cloud, a large number of stars form, having a value of [Na/Fe] very
skewed toward the value of the large cloud; for this reason
$N$([Na/Fe]) shows a clear bimodal structure with two peaks close to
the edges of the [Na/Fe] range of values. 

The vertical dashed lines in
Fig.  \ref{fig:distributions} indicate the value [Na/Fe]=0.4 which is
adopted by \citet{carr10} to separate the FG from the SG (see section
\ref{sec:fgmas}). The figure clearly shows that, for $x=0.05$,
inhomogeneous mixing can not populate the SG region ([Na/Fe]$>0.4$) for
any choice of $\alpha$. This conclusion does not change even if we
relax some of the other assumptions described above (and which play a
minor role in shaping $N$([Na/Fe]).

\begin{figure}    
\centering{\includegraphics[width=8cm]{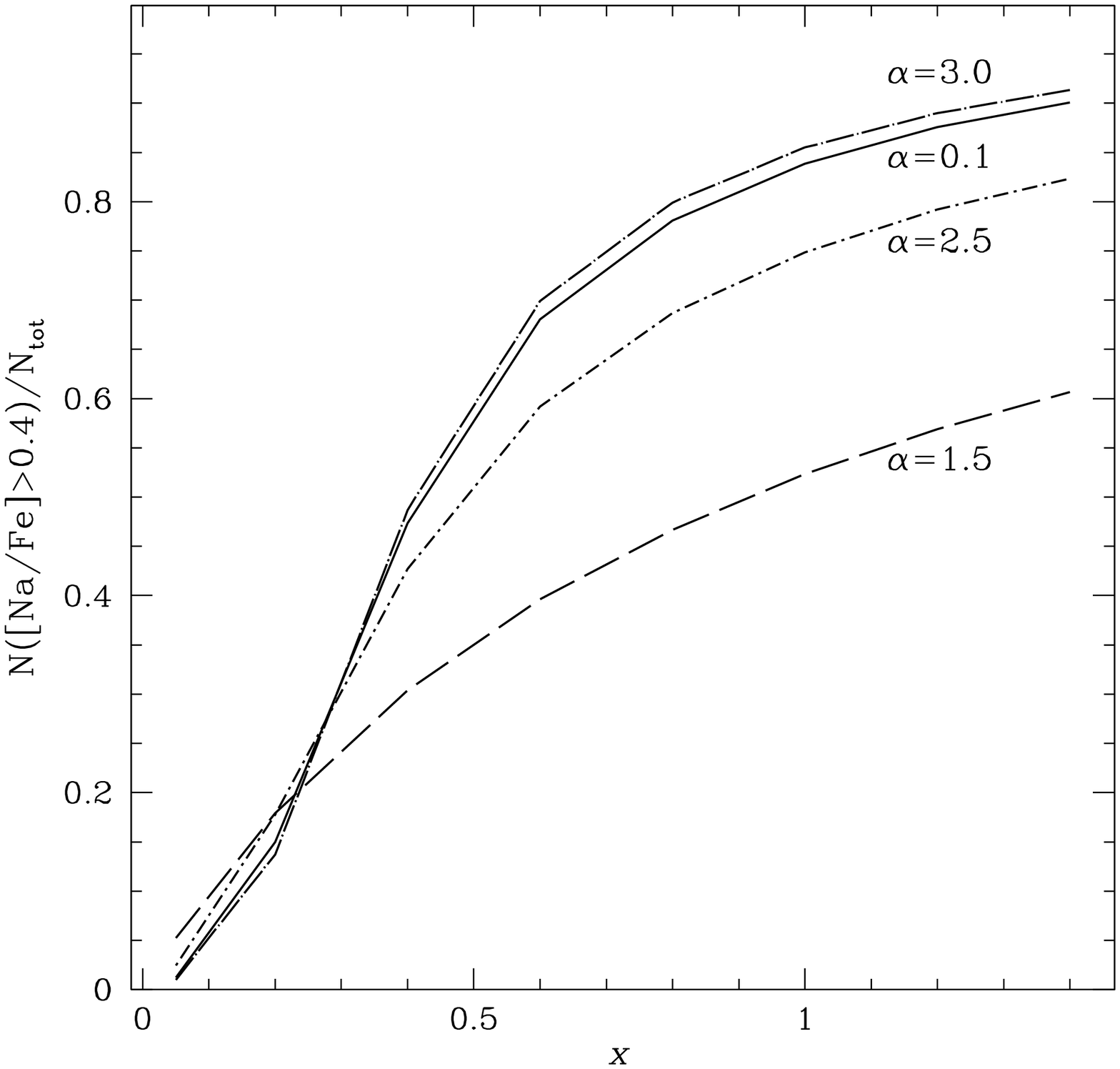}}
\caption{Fraction $\zeta$ of SG stars with [Na/Fe]$>0.4$ obtained by
  inhomogeneous mixing as function of $x$, and for different values of
  $\alpha$.  }
\label{fig:list} 
\end{figure}

We investigate now the influence of a possible inhomogeneous mixing
models with $x\sim 1$, that is, in models in which the initial FG mass
was much higher than today. Figure \ref{fig:list} illustrates the
fraction $\zeta$ of SG stars with [Na/Fe]$>0.4$ present in
$N$([Na/Fe]) as a function of $x$, for different values of $\alpha$. It
is clearly shown that, for large values of $x$, the fraction of new
stars really appearing as SG stars to an observer oscillates in a range
0.6-0.9,depending on $\alpha$. 

From equation \ref{eq:ratio} (see text),
the apparent ratio between the two populations is

\begin{equation}
\label{eq:ratios}
0<{M_{\rm SG} \over M_{\rm FG}}<\xi 
\end{equation}
\noindent
depending on the value of $\zeta$. Note that the maximum ratio
is obtained for $\zeta=1$, that is for the homogeneous mixing, when
all the SG stars have [Na/Fe]$>0.4$ (see the vertical solid lines in
the left panels of Fig. \ref{fig:distributions}).

Assuming $\xi=0.05$
(see section \ref{sec:fgmas}), it turns out from equation
\ref{eq:ratio} that the apparent ratio between the masses of the two
populations is a fraction $\sim \zeta$ of the true value (holding in
the homogeneous case). As a consequence, models with inhomogeneous
mixing need the initial FG mass to be increased by a factor 1.1-1.7
with respect to homogeneous models to obtain the same amount of
recognizable $M_{\rm SG}$.

\label{lastpage}

\end{document}